\documentclass[aps,prl,reprint,superscriptaddress,amsmath,amssymb]{revtex4-2}

\usepackage[T1]{fontenc}
\usepackage{lmodern}

\usepackage{amsmath,amssymb}
\usepackage{graphicx}
\usepackage{bm}
\usepackage{physics}   % optional; remove if you dislike its macros
\usepackage{siunitx}
\usepackage{hyperref}
\usepackage{xcolor}
\begin{document}
%\mciteErrorOnUnknownfalse
% ---------- metadata ----------
\title{Correlation-Enabled Beatings in Two-Dimensional Electronic Spectroscopy}

\author{Sirui Chen}
\email{schen931@gatech.edu}
\affiliation{School of Physics, Georgia Institute of Technology, Atlanta, Georgia 30332, United States}

%\author{Lili Wang}
%\affiliation{Department of Chemistry, Emory University,
%Atlanta, Georgia 30322, United States}

\author{Dragomir Davidović}
\affiliation{School of Physics, Georgia Institute of Technology, Atlanta, Georgia 30332, United States}

% ---------- Abstract (intentionally omitted for now) ----------
% \begin{abstract}
% \end{abstract}

\begin{abstract}
Long-lived beatings in two-dimensional electronic spectroscopy (2DES) remain difficult to interpret within standard excitonic open-system models, which typically assume factorized initialization and predict rapid coherence decay. We show that persistent beatings can arise from a correlation-driven mechanism that requires both slow bath memory and ultrafast pulse sequences that propagate system--bath correlations across optical interactions. In this regime, the pulse sequence unitarily dresses the bath-memory contribution and activates nonsecular population--coherence transfer during field-free evolution, sustaining coherence signatures far beyond factorized or weak-memory descriptions. Rather than addressing what is oscillating (excitonic versus vibronic) or quantum--versus--classical semantics, this work reframes long-lived beatings as a protocol-level dynamical effect: correlation-mediated retrieval under ultrafast control.
\end{abstract}

\maketitle
%\section{Introduction}

Two-dimensional electronic spectroscopy (2DES) uses sequences of femtosecond pulses to map a third-order nonlinear response onto a two-frequency spectrum with distinct excitation and detection axes, providing a direct window into coherent dynamics, energy-transfer pathways, and environment-induced line-shape evolution in complex molecular systems~\cite{Jonas2003}.
Because the detected signal is generated by a controlled sequence of ultrafast light--matter interactions, 2DES is also an operational probe of open-quantum-system dynamics under external driving~\cite{Pisliakov2006,Mancal2006}.

A prominent example is the observation of beating structure in 2DES of photosynthetic pigment--protein complexes and related multichromophoric aggregates~\cite{Engel2007}.
These oscillations were initially discussed as persistent electronic coherences, motivating extensive efforts to connect coherent dynamics with energy transport~\cite{Engel2007,Panitchayangkoon2010}.
Subsequent work established that vibrational and vibronic pathways can generate qualitatively similar oscillatory features in the same observables, particularly when vibrational modes are near-resonant with excitonic energy gaps and selected Liouville pathways enhance ground-state contributions~\cite{Thyrhaug2018,Plenio2013,Christensson2012}.
Accordingly, beating observations alone do not uniquely determine whether the underlying dynamics should be labeled ``electronic'' or ``vibronic,'' nor what is operationally ``quantum'' about the signal~\cite{Duan2017,Biswas2022}. Instead, we reframe long-lived beatings as a protocol-level question: do they reflect retention of system coherence, or correlation-mediated retrieval enabled by the pulse sequence?

The protocol-level ingredient typically absent from conventional models is this: when bath correlations persist over inter-pulse delays and waiting times, ultrafast control transfers pre-existing system--bath correlations across pulse boundaries, reshaping the waiting-time dynamics probed by 2DES.
This connects directly to our recent work on correlated open quantum systems under fast control operations~\cite{Chen2025}, where we showed that long bath memory, relative to gate times, enables control to transport system--bath correlations across the operation and thereby modify subsequent reduced dynamics. 
In this regime the reduced evolution is naturally described by two dynamical primitives (or ``two maps''): a correlation-dressed primitive associated with equilibration and memory, and a post-operation primitive that does not encode the same correlation content.
This viewpoint is consistent with and extends the general open-system fact that pre-existing correlations can invalidate single-map descriptions and make control outcomes depend on preparation history~\cite{Paz2019}.

Recent years have seen much progress in non-Markovian simulation methods for excitonic energy transfer and ultrafast spectroscopy, including microscopic simulations that reproduce experimentally relevant beating structures in realistic pigment--protein models and emphasize the importance of structured environments~\cite{Lorenzoni2025}.
However, in standard pulsed-spectroscopy workflows these approaches are typically deployed with a factorized initialization at the first-pulse boundary,
$\varrho_{\mathrm{tot}}(0)=\varrho_{S}(0)\otimes\varrho_{B}^{\mathrm{th}}$,
so correlation transfer across the pulse sequence is excluded by construction.

\begin{figure*}[t]
\centering
\includegraphics[width=0.75\linewidth]{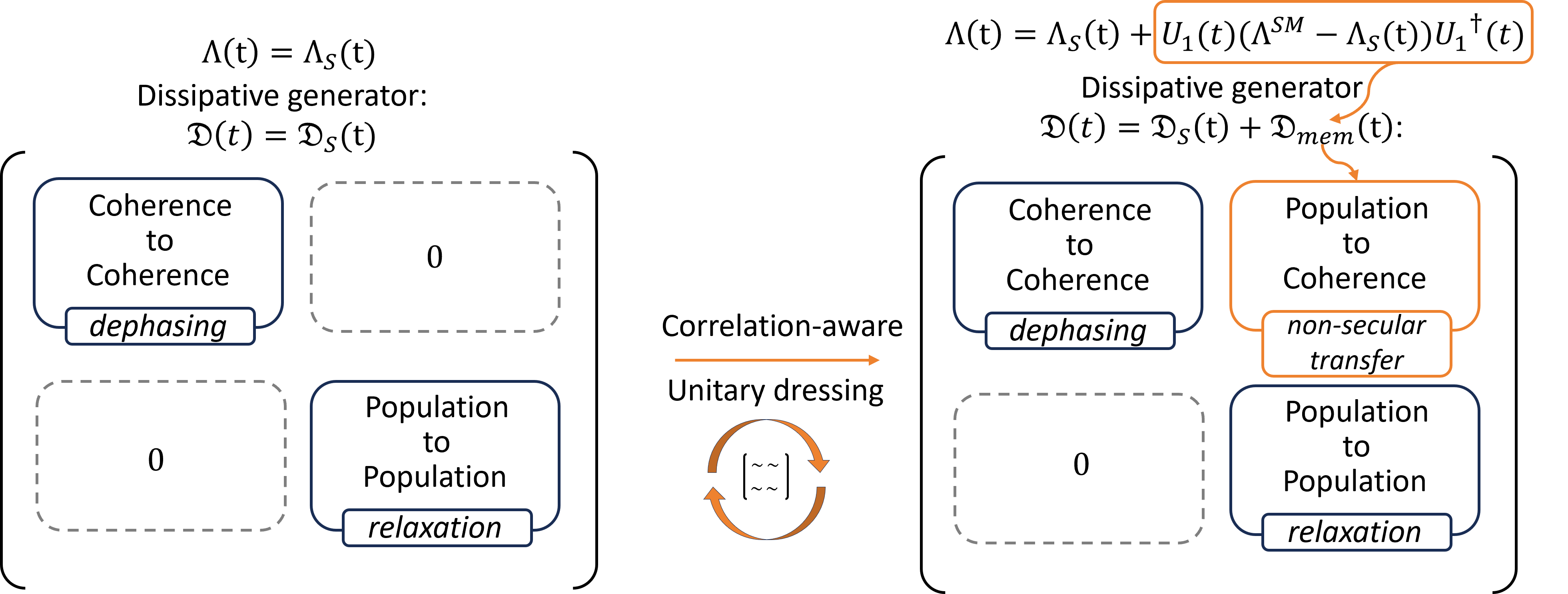}
\caption{Schematic: factorized single-map dynamics yields only population--population and coherence--coherence channels ($\mathcal{D}_S$), while pulse-dressed bath memory adds $\mathcal{D}_{\mathrm{mem}}$ and activates nonsecular population--coherence transfer. In either the weak-memory or factorized limit, $\mathcal{D}_{\mathrm{mem}}\!\approx\!0$.}
\label{fig:scheme_nonsecular}
\end{figure*}

Within our correlated open-system framework, vibronic oscillations do not by themselves identify the mechanism underlying persistent beatings.
Discrete vibrational structure and non-flat low-frequency spectral densities can both extend bath memory and enable population--coherence transfer during waiting-time evolution.
Recent microscopic simulations find long-lived signatures without fine-tuned vibron--exciton resonance~\cite{Lorenzoni2025}.
Instead, our mechanism relies on a broader premise: environmental correlations with long memory are prevalent on ultrafast timescales, and the pulse sequence can transport and unitarily dress pre-existing system--bath correlations.

To demonstrate these points in a spectroscopic setting, we construct a correlation-aware framework based on dynamical state preparation combined with time-dependent Bloch--Redfield dynamics, which explicitly tracks the transfer of system--bath correlations under ultrafast driving~\cite{Chen2025}.
Bloch--Redfield theory contains the essential ingredient required to capture correlation transfer across pulses: population-to-coherence transfer is accurate to leading order in the weak coupling constant, ensuring that coherences are computed with controlled precision~\cite{Thingna2012,Fleming2011,Tupkary2022,Crowder2024}.

We simulate rephasing and nonrephasing third-order signals for minimal excitonic models and find that long-lived beating signatures emerge in the memory-bath regime.
This yields a unified interpretation of persistent 2DES beatings as an open-system dynamical effect driven by the joint action of ultrafast control and bath memory, independent of whether the oscillatory features are labeled electronic, vibrational, or vibronic.

%\section{Correlation-aware dynamics under pulsed driving}
%\label{sec:dynamics_main}

A three-pulse 2DES protocol applies a prescribed sequence of ultrafast operations to an open quantum system.
When bath correlations persist over the inter-pulse delays and the waiting time, the reduced dynamics in a given segment can depend explicitly on correlations established in earlier segments.
To capture this regime explicitly, we do \emph{not} reset the system--bath state to a factorized form at pulse boundaries; instead, we retain \textbf{pre-existing system--bath correlations} that survive each pulse and modify the subsequent relaxation and dephasing dynamics~\cite{Chen2025}.

We describe the driven reduced system dynamics in the rotating frame using a time-dependent Bloch--Redfield generator without secular approximation,
\begin{equation}
\frac{d}{dt}\varrho(t)=\mathcal{D}(t)\,\varrho(t),
\label{eq:ME_I}
\end{equation}
and treat each ultrafast optical pulse as an explicit unitary operation on the system.
The details of the open quantum system model are provided in Appendix~A.
The key distinction from conventional treatments is that bath memory generates an additional, history-dependent contribution to the dissipator that carries information about prior system--bath correlations.

For clarity, we decompose the generator as
\begin{equation}
\mathcal{D}(t)=\mathcal{D}_{S}(t)+\mathcal{D}_{\mathrm{mem}}(t),
\label{eq:D_split_main}
\end{equation}
where $\mathcal{D}_{S}(t)$ is the conventional time-dependent generator used in standard factorized-initial-condition treatments, and
$\mathcal{D}_{\mathrm{mem}}(t)$ encodes the influence of \textbf{pre-existing system--bath correlations} carried forward across the pulse sequence~\cite{Chen2025}.
The detailed derivations are provided in Appendix~A and Appendix~B.

Figure~\ref{fig:scheme_nonsecular} summarizes the central conceptual distinction of this work and makes clear that the persistence of beating structure is controlled not by its microscopic origin, but by whether pulse sequences can retrieve coherence from system–bath correlations.
In conventional factorized treatments (left), the dissipator $\mathcal{D}_{S}(t)$ contains only coherence–coherence and population–population channels, whereas in the correlation-aware framework (right) a pulse-dressed memory contribution $\mathcal{D}_{\mathrm{mem}}(t)$ activates nonsecular population–coherence transfer.

Crucially, the difference between the two panels is not the bath itself but whether the pulse sequence is allowed to act on pre-existing system–bath correlations, thereby changing which dissipative operator sectors survive coarse graining over the excitonic splitting.
$\mathcal{D}_{\mathrm{mem}}(t)$ is \emph{unitarily dressed} at the pulse boundary: the pulse unitary $U_j$ conjugates the memory contribution, $\,\mathcal{D}_{\mathrm{mem}}\mapsto U_j\,\mathcal{D}_{\mathrm{mem}}\,U_j^\dagger$, thereby rotating bath memory operators in Liouville space. 
The resulting \emph{two-map} dynamics reshapes dissipative pathways and can activate \textbf{nonsecular population--coherence transfer} at rates comparable to dephasing, yielding coherence signatures inaccessible within single-map secular descriptions.

The magnitude and relevance of $\mathcal{D}_{\mathrm{mem}}(t)$ are controlled by the bath memory time.
When bath correlations decay rapidly (memoryless or weak-memory regimes), $\mathcal{D}_{\mathrm{mem}}(t)\approx 0$ and the dynamics reduces to $\mathcal{D}(t)\approx \mathcal{D}_{S}(t)$, so the protocol history does not introduce additional dynamical channels.
In contrast, when bath correlations persist over the inter-pulse delays and waiting time, $\mathcal{D}_{\mathrm{mem}}(t)$ remains appreciable, and the pulse sequence can leverage its unitary dressing to activate nonsecular population--coherence pathways that are absent in factorized or single-map treatments.

In three-pulse 2DES, the measured field corresponds to the leading \emph{third-order} response of an open quantum system to a prescribed sequence of ultrafast light--matter interactions~\cite{Biswas2022,Chen2010}.
Because our correlation-aware field-free evolution allows nonsecular population--coherence transfer during the waiting time, standard pathway-based response-function constructions that propagate optical-action superoperators $V\rho=[\mu,\rho]$ can obscure the relevant dynamics~\cite{Chen2010,Biswas2022}.
We therefore extract the third-order contribution by inclusion--exclusion from the full driven evolution; details of this procedure are provided in Appendix~C.

We simulate the FMO site-1/site-3 dimer with $\epsilon_A=12410~\mathrm{cm^{-1}}$, $\epsilon_B=12210~\mathrm{cm^{-1}}$, $J=5.5~\mathrm{cm^{-1}}$, dipoles $\mu_A=1.0$ and $\mu_B=-0.8$, and a thermal initial state at $77~\mathrm{K}$~\cite{Adolphs2006,Braver2021,Nalbach2011}.
The bath cutoff is $\omega_c=(100~\mathrm{fs})^{-1}$.
Between pulses, the interaction-picture master equation is integrated using a fourth-order Runge--Kutta scheme.

Long-lived beating structure requires (i) slow bath memory and (ii) correlation-aware propagation that pulse-dresses the memory term to activate nonsecular population--coherence transfer.
This mechanism provides an excitonic route to reconcile longstanding discrepancies between experiment and theory.
In 2DES measurements of FMO complexes, beatings persisting beyond $\sim$600~fs have been reported, whereas coherence in many standard excitonic open-system treatments decays on $\sim$200~fs timescales~\cite{Engel2007,Panitchayangkoon2010,Ishizaki2009}.

To test these requirements in a controlled manner, we simulate three-pulse 2DES for the FMO site-1/site-3 dimer and compare three dynamical settings that differ only in
(a) the bath memory and
(b) whether pre-existing system--bath correlations are retained across pulse boundaries.
In all cases, the third-order signal $S^{(3)}$ is extracted from the full driven open-system dynamics using the operational inclusion--exclusion procedure described above, ensuring that nonsecular effects are retained.

\begin{figure}[htbp]
  \centering
  \colorbox{white}{\includegraphics[width=\linewidth]{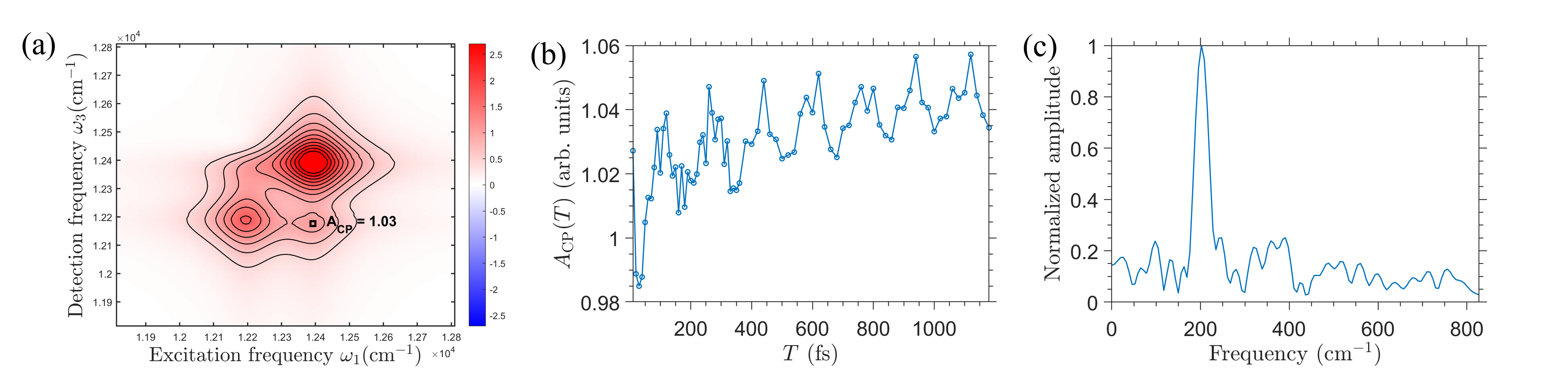}}
  \caption{
  \textbf{Sub-Ohmic slow bath ($s=0.9$) with correlation-aware dynamics.}
  (a) Absorptive 2D spectrum at $T=10$~fs; the marked cross peak defines the amplitude observable $A_{\mathrm{CP}}$.
  (b) Waiting-time cross-peak amplitude $A_{\mathrm{CP}}(T)$ exhibiting persistent oscillations.
  (c) Fourier spectrum of $A_{\mathrm{CP}}(T)$ showing a dominant, spectrally concentrated beating component.
  }
  \label{fig:slowbath_CA}
\end{figure}

The following comparisons isolate the operational distinction between coherence retention within a single dynamical map and correlation-mediated retrieval enabled by pulse-dressed bath memory. Figure~\ref{fig:slowbath_CA} summarizes the sub-Ohmic slow-bath case ($s=0.9$) under correlation-aware propagation.
Panel (a) shows an absorptive 2D spectrum snapshot at a short waiting time, $T=10$~fs, where diagonal and off-diagonal features are well resolved.
We quantify the waiting-time coherence by tracking a representative cross-peak amplitude $A_{\mathrm{CP}}(T)$.
As shown in panel (b), $A_{\mathrm{CP}}(T)$ exhibits pronounced oscillations that persist into the picosecond regime.
The corresponding Fourier spectrum in panel (c) displays a spectrally concentrated beating component.
The dominant peak near $200~\mathrm{cm}^{-1}$ aligns with the excitonic energy splitting.

Such persistent beatings in FMO 2DES experiments have been discussed extensively, with interpretations ranging from electronic to vibronic coherence~\cite{Engel2007,Panitchayangkoon2010,Christensson2012,Tiwari2012,Milota2013}.
Within our correlated open-system framework, these beatings arise from a dynamical mechanism: slow bath memory preserves nonsecular population--coherence transfer over inter-pulse delays and the waiting time, allowing the pulse sequence to convert system--bath correlations carried from earlier segments into post-pulse coherence dynamics.

\begin{figure}[t]
  \centering
  \colorbox{white}{\includegraphics[width=\linewidth]{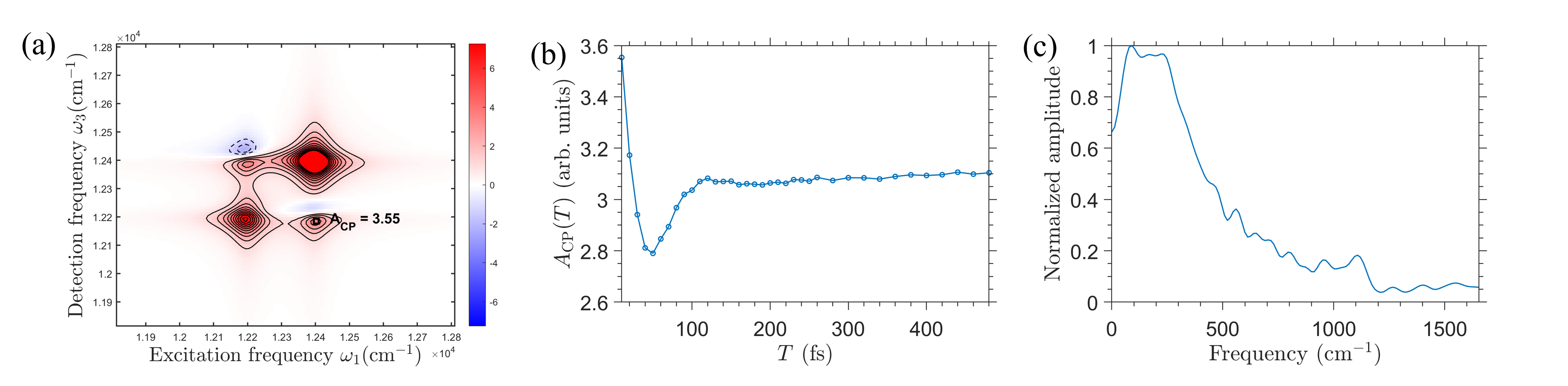}}
  \caption{
  \textbf{Ohmic bath ($s=1$) with correlation-aware dynamics (weak-memory case).}
  (a) Absorptive 2D spectrum at $T=10$~fs.
  (b) $A_{\mathrm{CP}}(T)$ showing strongly reduced  oscillatory behavior relative to Fig.~\ref{fig:slowbath_CA}.
  (c) Fourier spectrum of $A_{\mathrm{CP}}(T)$ without a comparably sharp peak indicating a persistent beating component.
  }
  \label{fig:ohmic_CA}
\end{figure}

We next switch to an Ohmic environment ($s=1$), which corresponds to more rapidly decaying bath correlations and thus a shorter memory time.
Figure~\ref{fig:ohmic_CA} shows that while the early-time 2D spectral features at $T=10$~fs remain qualitatively similar (panel a), the waiting-time trace (panel b) exhibits reduced oscillatory structure.
The Fourier spectrum (panel c) correspondingly lacks a comparably sharp beating component.
This trend is consistent with the expectation that when bath correlations decay rapidly, the memory contribution becomes negligible and the dynamics reduces to a conventional form in which population and coherence sectors are effectively decoupled during field-free evolution~\cite{Pisliakov2006,Mancal2006,Biswas2022}.
In this weak-memory regime, $\mathcal{D}_{\mathrm{mem}}(t)\to 0$ quickly, so pulse-induced dressing has limited leverage to reshape dissipative pathways.

\begin{figure}[t]
  \centering
  \colorbox{white}{\includegraphics[width=\linewidth]{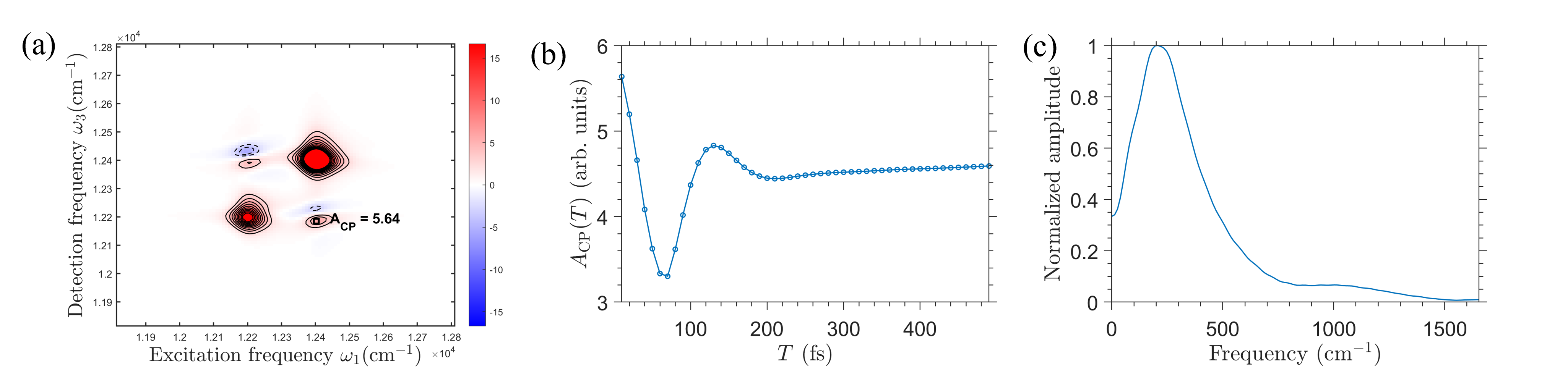}}
  \caption{
  \textbf{Sub-Ohmic slow bath ($s=0.9$) \emph{without} correlation-aware propagation (factorized reset at pulse boundaries).}
  (a) Absorptive 2D spectrum at $T=10$~fs.
  (b) $A_{\mathrm{CP}}(T)$ shows strongly reduced oscillatory structure relative to the correlation-aware case, indicating that slow bath memory alone is insufficient without pulse-induced dressing of pre-existing system--bath correlations.
  (c) Fourier spectrum of $A_{\mathrm{CP}}(T)$ without a comparably sharp beating component.
  }
  \label{fig:slowbath_noCA}
\end{figure}

We keep the same sub-Ohmic slow bath ($s=0.9$) but remove correlation awareness by enforcing a factorized reset at pulse boundaries, corresponding to a conventional non-Markovian Bloch--Redfield treatment that does not carry pre-existing system--bath correlations across segments.
Figure~\ref{fig:slowbath_noCA} shows that this change qualitatively alters the waiting-time behavior: despite the slow bath, $A_{\mathrm{CP}}(T)$ no longer exhibits the pronounced oscillatory structure observed in Fig.~\ref{fig:slowbath_CA}, and the Fourier spectrum does not show a similarly prominent sharp component.
This contrast isolates the role of pre-existing system--bath correlations: bath memory alone is insufficient unless the dynamics retains correlation content through the pulse sequence, so that unitary dressing of the memory contribution can activate nonsecular population--coherence transfer.
Equivalently, the oscillatory features in our simulations are not simply a consequence of using a non-Markovian generator, but specifically of \emph{correlation transfer} enabled by correlation-aware propagation under pulsed driving.

Figs.~\ref{fig:slowbath_CA}--\ref{fig:slowbath_noCA} isolate correlation-mediated retrieval: beatings appear only when bath memory persists and correlations are propagated across pulses.

\begin{figure}[t]
\centering
\colorbox{white}{\includegraphics[width=0.98\linewidth]{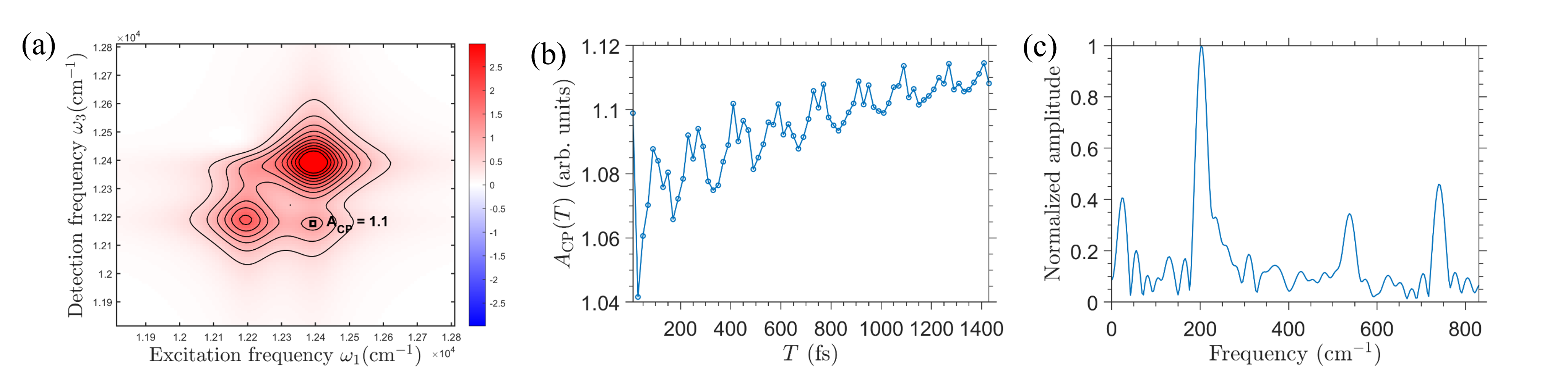}}
\caption{
\textbf{Structured low-frequency bath.}
Representative three-panel summary for the structured-bath model defined in Eqs.~\eqref{eq:J_structured_def}--\eqref{eq:J_lin_tail}, shown in the same format as the canonical spectral-density cases.
(a) Absorptive 2D spectrum at $T=10$~fs; the marked cross peak defines the amplitude observable $A_{\mathrm{CP}}$.
(b) Waiting-time cross-peak amplitude $A_{\mathrm{CP}}(T)$ exhibiting pronounced oscillatory structure.
(c) Fourier spectrum of $A_{\mathrm{CP}}(T)$ showing a dominant peak near $200~\mathrm{cm}^{-1}$, consistent with the excitonic energy splitting.
The re-emergence of beating structure reflects the restoration of slow bath memory via low-frequency spectral structure, rather than resonant vibronic matching.
}
\label{fig:structured_bath_triple}
\end{figure}

This correlation-driven route is compatible with vibronic discussions in the 2DES literature, because underdamped vibrational structure provides a natural microscopic source of slow bath memory~\cite{Christensson2012,Tiwari2012,Milota2013}.
To illustrate this connection (see Appendix~A\ref{app:model_class}), we add a structured low-frequency component to the bath while preserving an Ohmic form in the limit $\omega\to 0$.
We find that this structure prolongs bath correlations and keeps the memory term $\mathcal{D}_{\mathrm{mem}}(t)$ appreciable over inter-pulse delays and the waiting time, \emph{without} requiring resonant matching between the structure and the excitonic splitting.
Using the same three-panel diagnostic as in Figs.~\ref{fig:slowbath_CA}--\ref{fig:slowbath_noCA}, the enhanced memory appears as a slowly decaying oscillatory component in the cross-peak trace $A_{\mathrm{CP}}(T)$ (Fig.~\ref{fig:structured_bath_triple}).
These signatures closely parallel those of the canonical slow-memory bath, indicating that structured low-frequency modes can return the system to the regime in which \textbf{nonsecular population--coherence transfer} is activated by the pulse sequence.

Despite the presence of coherence signatures in this regime, we find no evidence of quantum entanglement in any of the simulations.
A detailed analysis is provided in Appendix~E.

In conclusion, long-lived beating structure in 2DES can emerge from a correlation-driven open-system mechanism that requires slow bath memory and the propagation of system–bath correlations across ultrafast pulses. In this two-map regime, the pulse sequence unitarily dresses bath memory and activates nonsecular population–coherence transfer during field-free evolution, producing coherence signatures that are absent in factorized or weak-memory, single-map descriptions. The central implication is that long-lived 2DES beatings should be interpreted not as evidence for a particular microscopic origin, but as signatures of correlation-mediated retrieval enabled by ultrafast control.

Vibronic interpretations are consistent with this picture: vibronic structure can provide long bath memory, but coherence activation does not require fine-tuned vibronic resonance~\cite{Lorenzoni2025}. Resonant pathways in cyanine-dye J-aggregates~\cite{Lim2015} are captured within factorized descriptions and may dominate when correlation-induced memory effects are weak. For complex pigment–protein systems, where correlated fluctuations have been proposed~\cite{Lee2007,Ishizaki2010}, low-frequency spectral structure naturally supports the long-memory conditions required for the two-map regime. Establishing a quantitative correspondence between correlation-transfer and correlated-noise descriptions is an important direction for future work. Capturing narrow, spectrally congested features of the spectral density may require nonperturbative methods beyond the present scope~\cite{Lorenzoni2025}.

We thank Jiahao Chen for fruitful discussions and Lili Wang for reviewing the manuscript and providing valuable feedback.
We acknowledge seed support from the School of Physics, Georgia Tech.

\newpage

\bibliography{ref}

@article{Paz2019,
  title = {Dynamics of initially correlated open quantum systems: Theory and applications},
  author = {Paz-Silva, Gerardo A. and Hall, Michael J. W. and Wiseman, Howard M.},
  journal = {Phys. Rev. A},
  volume = {100},
  issue = {4},
  pages = {042120},
  numpages = {19},
  year = {2019},
  month = {Oct},
  publisher = {American Physical Society},
  doi = {10.1103/PhysRevA.100.042120},
  url = {https://link.aps.org/doi/10.1103/PhysRevA.100.042120}
}

@Article{Jonas2003,
  author    = {Jonas, David M.},
  journal   = {Annu. Rev. Phys. Chem.},
  title     = {Two-Dimensional Femtosecond Spectroscopy},
  year      = {2003},
  issn      = {1545-1593},
  month     = oct,
  number    = {1},
  pages     = {425--463},
  volume    = {54},
  doi       = {10.1146/annurev.physchem.54.011002.103907},
  fjournal  = {Annual Review of Physical Chemistry},
  publisher = {Annual Reviews},
}

@Article{Engel2007,
  author    = {Engel, Gregory S. and Calhoun, Tessa R. and Read, Elizabeth L. and Ahn, Tae-Kyu and Mančal, Tomáš and Cheng, Yuan-Chung and Blankenship, Robert E. and Fleming, Graham R.},
  journal   = {Nature},
  title     = {Evidence for wavelike energy transfer through quantum coherence in photosynthetic systems},
  year      = {2007},
  issn      = {1476-4687},
  month     = apr,
  number    = {7137},
  pages     = {782--786},
  volume    = {446},
  doi       = {10.1038/nature05678},
  publisher = {Springer Science and Business Media LLC},
}

@Article{Panitchayangkoon2010,
  author    = {Panitchayangkoon, Gitt and Hayes, Dugan and Fransted, Kelly A. and Caram, Justin R. and Harel, Elad and Wen, Jianzhong and Blankenship, Robert E. and Engel, Gregory S.},
  journal   = {Proc. Natl. Acad. Sci.},
  title     = {Long-lived quantum coherence in photosynthetic complexes at physiological temperature},
  year      = {2010},
  issn      = {1091-6490},
  month     = jul,
  number    = {29},
  pages     = {12766--12770},
  volume    = {107},
  doi       = {10.1073/pnas.1005484107},
  fjournal  = {Proceedings of the National Academy of Sciences},
  publisher = {Proceedings of the National Academy of Sciences},
}

@Article{Tiwari2012,
  author    = {Tiwari, Vivek and Peters, William K. and Jonas, David M.},
  journal   = {Proc. Natl. Acad. Sci.},
  title     = {Electronic resonance with anticorrelated pigment vibrations drives photosynthetic energy transfer outside the adiabatic framework},
  year      = {2012},
  issn      = {1091-6490},
  month     = dec,
  number    = {4},
  pages     = {1203--1208},
  volume    = {110},
  doi       = {10.1073/pnas.1211157110},
  fjournal  = {Proceedings of the National Academy of Sciences},
  publisher = {Proceedings of the National Academy of Sciences},
}

@Article{Milota2013,
  author    = {Milota, Franz and Prokhorenko, Valentyn I. and Mancal, Tomas and von Berlepsch, Hans and Bixner, Oliver and Kauffmann, Harald F. and Hauer, Jürgen},
  journal   = {J. Phys. Chem. A},
  title     = {Vibronic and Vibrational Coherences in Two-Dimensional Electronic Spectra of Supramolecular J-Aggregates},
  year      = {2013},
  issn      = {1520-5215},
  month     = mar,
  number    = {29},
  pages     = {6007--6014},
  volume    = {117},
  doi       = {10.1021/jp3119605},
  fjournal  = {The Journal of Physical Chemistry A},
  publisher = {American Chemical Society (ACS)},
}

@Article{Thyrhaug2018,
  author    = {Thyrhaug, Erling and Tempelaar, Roel and Alcocer, Marcelo J. P. and Žídek, Karel and Bína, David and Knoester, Jasper and Jansen, Thomas L. C. and Zigmantas, Donatas},
  journal   = {Nat. Chem.},
  title     = {Identification and characterization of diverse coherences in the Fenna–Matthews–Olson complex},
  year      = {2018},
  issn      = {1755-4349},
  month     = may,
  number    = {7},
  pages     = {780--786},
  volume    = {10},
  doi       = {10.1038/s41557-018-0060-5},
  fjournal  = {Nature Chemistry},
  publisher = {Springer Science and Business Media LLC},
}

@Article{Duan2017,
  author    = {Duan, Hong-Guang and Prokhorenko, Valentyn I. and Cogdell, Richard J. and Ashraf, Khuram and Stevens, Amy L. and Thorwart, Michael and Miller, R. J. Dwayne},
  journal   = {Proc. Natl. Acad. Sci.},
  title     = {Nature does not rely on long-lived electronic quantum coherence for photosynthetic energy transfer},
  year      = {2017},
  issn      = {1091-6490},
  month     = jul,
  number    = {32},
  pages     = {8493--8498},
  volume    = {114},
  doi       = {10.1073/pnas.1702261114},
  fjournal  = {Proceedings of the National Academy of Sciences},
  publisher = {Proceedings of the National Academy of Sciences},
}

@Article{Chen2025,
  author    = {Chen, Sirui and Chen, Jiahao and Davidović, Dragomir},
  journal   = {Phys. Rev. Applied},
  title     = {Gate-based initialization and fidelity in correlated open quantum systems},
  year      = {2025},
  issn      = {2331-7019},
  month     = sep,
  number    = {3},
  volume    = {24},
  doi       = {10.1103/gbmy-tp3g},
  fjournal  = {Physical Review Applied},
  publisher = {American Physical Society (APS)},
}

@Article{Pisliakov2006,
  author    = {Pisliakov, Andrei V. and Mančal, Tomáš and Fleming, Graham R.},
  journal   = {J. Chem. Phys.},
  title     = {Two-dimensional optical three-pulse photon echo spectroscopy. II. Signatures of coherent electronic motion and exciton population transfer in dimer two-dimensional spectra},
  year      = {2006},
  issn      = {1089-7690},
  month     = jun,
  number    = {23},
  volume    = {124},
  doi       = {10.1063/1.2200705},
  fjournal  = {The Journal of Chemical Physics},
  publisher = {AIP Publishing},
}

@Article{Mancal2006,
  author    = {Mančal, Tomáš and Pisliakov, Andrei V. and Fleming, Graham R.},
  journal   = {J. Chem. Phys.},
  title     = {Two-dimensional optical three-pulse photon echo spectroscopy. I. Nonperturbative approach to the calculation of spectra},
  year      = {2006},
  issn      = {1089-7690},
  month     = jun,
  number    = {23},
  volume    = {124},
  doi       = {10.1063/1.2200704},
  fjournal  = {The Journal of Chemical Physics},
  publisher = {AIP Publishing},
}

@Article{Ishizaki2009,
  author    = {Ishizaki, Akihito and Fleming, Graham R.},
  journal   = {J. Chem. Phys.},
  title     = {Unified treatment of quantum coherent and incoherent hopping dynamics in electronic energy transfer: Reduced hierarchy equation approach},
  year      = {2009},
  issn      = {1089-7690},
  month     = jun,
  number    = {23},
  volume    = {130},
  doi       = {10.1063/1.3155372},
  fjournal  = {The Journal of Chemical Physics},
  publisher = {AIP Publishing},
}

@Article{Chen2010,
  author    = {Chen, Liping and Zheng, Renhui and Shi, Qiang and Yan, YiJing},
  journal   = {J. Chem. Phys.},
  title     = {Two-dimensional electronic spectra from the hierarchical equations of motion method: Application to model dimers},
  year      = {2010},
  issn      = {1089-7690},
  month     = jan,
  number    = {2},
  volume    = {132},
  doi       = {10.1063/1.3293039},
  fjournal  = {The Journal of Chemical Physics},
  publisher = {AIP Publishing},
}

@Article{Biswas2022,
  author    = {Biswas, Somnath and Kim, JunWoo and Zhang, Xinzi and Scholes, Gregory D.},
  journal   = {Chem. Rev.},
  title     = {Coherent Two-Dimensional and Broadband Electronic Spectroscopies},
  year      = {2022},
  issn      = {1520-6890},
  month     = jan,
  number    = {3},
  pages     = {4257--4321},
  volume    = {122},
  doi       = {10.1021/acs.chemrev.1c00623},
  fjournal  = {Chemical Reviews},
  publisher = {American Chemical Society (ACS)},
}

@Article{Plenio2013,
  author    = {Plenio, M. B. and Almeida, J. and Huelga, S. F.},
  journal   = {J. Chem. Phys.},
  title     = {Origin of long-lived oscillations in 2D-spectra of a quantum vibronic model: Electronic versus vibrational coherence},
  year      = {2013},
  issn      = {1089-7690},
  month     = dec,
  number    = {23},
  volume    = {139},
  doi       = {10.1063/1.4846275},
  fjournal  = {The Journal of Chemical Physics},
  publisher = {AIP Publishing},
}

@Article{Christensson2012,
  author    = {Christensson, Niklas and Kauffmann, Harald F. and Pullerits, Tõnu and Mančal, Tomáš},
  journal   = {J. Phys. Chem. B},
  title     = {Origin of Long-Lived Coherences in Light-Harvesting Complexes},
  year      = {2012},
  issn      = {1520-5207},
  month     = jun,
  number    = {25},
  pages     = {7449--7454},
  volume    = {116},
  doi       = {10.1021/jp304649c},
  publisher = {American Chemical Society (ACS)},
}

@Article{Ishizaki2010,
  author    = {Ishizaki, Akihito and Calhoun, Tessa R. and Schlau-Cohen, Gabriela S. and Fleming, Graham R.},
  journal   = {Phys. Chem. Chem. Phys.},
  title     = {Quantum coherence and its interplay with protein environments in photosynthetic electronic energy transfer},
  year      = {2010},
  issn      = {1463-9084},
  number    = {27},
  pages     = {7319},
  volume    = {12},
  doi       = {10.1039/c003389h},
  fjournal  = {Physical Chemistry Chemical Physics},
  publisher = {Royal Society of Chemistry (RSC)},
}

@Article{Thingna2012,
  author    = {Thingna, Juzar and Wang, Jian-Sheng and Hänggi, Peter},
  journal   = {J. Chem. Phys.},
  title     = {Generalized Gibbs state with modified Redfield solution: Exact agreement up to second order},
  year      = {2012},
  issn      = {1089-7690},
  month     = may,
  number    = {19},
  volume    = {136},
  doi       = {10.1063/1.4718706},
  fjournal  = {The Journal of Chemical Physics},
  publisher = {AIP Publishing},
}

@Article{Fleming2011,
  author    = {Fleming, C. H. and Cummings, N. I.},
  journal   = {Phys. Rev. E},
  title     = {Accuracy of perturbative master equations},
  year      = {2011},
  issn      = {1550-2376},
  month     = mar,
  number    = {3},
  pages     = {031117},
  volume    = {83},
  doi       = {10.1103/physreve.83.031117},
  fjournal  = {Physical Review E},
  publisher = {American Physical Society (APS)},
}

@Article{Tupkary2022,
  author    = {Tupkary, Devashish and Dhar, Abhishek and Kulkarni, Manas and Purkayastha, Archak},
  journal   = {Phys. Rev. A},
  title     = {Fundamental limitations in Lindblad descriptions of systems weakly coupled to baths},
  year      = {2022},
  issn      = {2469-9934},
  month     = mar,
  number    = {3},
  pages     = {032208},
  volume    = {105},
  doi       = {10.1103/physreva.105.032208},
  fjournal  = {Physical Review A},
  publisher = {American Physical Society (APS)},
}

@Article{Crowder2024,
  author    = {Crowder, Elyana and Lampert, Lance and Manchanda, Grihith and Shoffeitt, Brian and Gadamsetty, Srikar and Pei, Yiting and Chaudhary, Shantanu and Davidović, Dragomir},
  journal   = {Phys. Rev. A},
  title     = {Invalidation of the Bloch-Redfield equation in the sub-Ohmic regime via a practical time-convolutionless fourth-order master equation},
  year      = {2024},
  issn      = {2469-9934},
  month     = may,
  number    = {5},
  pages     = {052205},
  volume    = {109},
  doi       = {10.1103/physreva.109.052205},
  fjournal  = {Physical Review A},
  publisher = {American Physical Society (APS)},
}

@Article{Horodecki1996,
  author    = {Horodecki, Michał and Horodecki, Paweł and Horodecki, Ryszard},
  journal   = {Phys. Lett. A},
  title     = {Separability of mixed states: necessary and sufficient conditions},
  year      = {1996},
  issn      = {0375-9601},
  month     = nov,
  number    = {1–2},
  pages     = {1--8},
  volume    = {223},
  doi       = {10.1016/s0375-9601(96)00706-2},
  fjournal  = {Physics Letters A},
  publisher = {Elsevier BV},
}

@Article{Peres1996,
  author    = {Peres, Asher},
  journal   = {Phys. Rev. Lett.},
  title     = {Separability Criterion for Density Matrices},
  year      = {1996},
  issn      = {1079-7114},
  month     = aug,
  number    = {8},
  pages     = {1413--1415},
  volume    = {77},
  doi       = {10.1103/physrevlett.77.1413},
  fjournal  = {Physical Review Letters},
  publisher = {American Physical Society (APS)},
}

@Article{Renger2012,
  author    = {Renger, Thomas and Klinger, Alexander and Steinecker, Florian and Schmidt am Busch, Marcel and Numata, Jorge and Müh, Frank},
  journal   = {J. Phys. Chem. B},
  title     = {Normal Mode Analysis of the Spectral Density of the Fenna–Matthews–Olson Light-Harvesting Protein: How the Protein Dissipates the Excess Energy of Excitons},
  year      = {2012},
  issn      = {1520-5207},
  month     = dec,
  number    = {50},
  pages     = {14565--14580},
  volume    = {116},
  doi       = {10.1021/jp3094935},
  publisher = {American Chemical Society (ACS)},
}

@Article{Klinger2020,
  author    = {Klinger, Alexander and Lindorfer, Dominik and Müh, Frank and Renger, Thomas},
  journal   = {J. Chem. Phys.},
  title     = {Normal mode analysis of spectral density of FMO trimers: Intra- and intermonomer energy transfer},
  year      = {2020},
  issn      = {1089-7690},
  month     = dec,
  number    = {21},
  volume    = {153},
  doi       = {10.1063/5.0027994},
  fjournal  = {The Journal of Chemical Physics},
  publisher = {AIP Publishing},
}

@Article{Adolphs2006,
  author    = {Adolphs, Julia and Renger, Thomas},
  journal   = {Biophys. J.},
  title     = {How Proteins Trigger Excitation Energy Transfer in the FMO Complex of Green Sulfur Bacteria},
  year      = {2006},
  issn      = {0006-3495},
  month     = oct,
  number    = {8},
  pages     = {2778--2797},
  volume    = {91},
  doi       = {10.1529/biophysj.105.079483},
  fjournal  = {Biophysical Journal},
  publisher = {Elsevier BV},
}

@Article{Braver2021,
  author    = {Braver, Yakov and Valkunas, Leonas and Gelzinis, Andrius},
  journal   = {J. Chem. Theory Comput.},
  title     = {Quantum–Classical Approach for Calculations of Absorption and Fluorescence: Principles and Applications},
  year      = {2021},
  issn      = {1549-9626},
  month     = oct,
  number    = {11},
  pages     = {7157--7168},
  volume    = {17},
  doi       = {10.1021/acs.jctc.1c00777},
  fjournal  = {Journal of Chemical Theory and Computation},
  publisher = {American Chemical Society (ACS)},
}

@Article{Nalbach2011,
  author    = {Nalbach, P. and Braun, D. and Thorwart, M.},
  journal   = {Phys. Rev. E},
  title     = {Exciton transfer dynamics and quantumness of energy transfer in the Fenna-Matthews-Olson complex},
  year      = {2011},
  issn      = {1550-2376},
  month     = oct,
  number    = {4},
  pages     = {041926},
  volume    = {84},
  doi       = {10.1103/physreve.84.041926},
  fjournal  = {Physical Review E},
  publisher = {American Physical Society (APS)},
}

@Article{Lim2015,
  author    = {Lim, James and Paleček, David and Caycedo-Soler, Felipe and Lincoln, Craig N. and Prior, Javier and von Berlepsch, Hans and Huelga, Susana F. and Plenio, Martin B. and Zigmantas, Donatas and Hauer, Jürgen},
  journal   = {Nat. Commun.},
  title     = {Vibronic origin of long-lived coherence in an artificial molecular light harvester},
  year      = {2015},
  issn      = {2041-1723},
  month     = jul,
  number    = {1},
  volume    = {6},
  doi       = {10.1038/ncomms8755},
  fjournal  = {Nature Communications},
  publisher = {Springer Science and Business Media LLC},
}

@Article{Lee2007,
  author    = {Lee, Hohjai and Cheng, Yuan-Chung and Fleming, Graham R.},
  journal   = {Science},
  title     = {Coherence Dynamics in Photosynthesis: Protein Protection of Excitonic Coherence},
  year      = {2007},
  issn      = {1095-9203},
  month     = jun,
  number    = {5830},
  pages     = {1462--1465},
  volume    = {316},
  doi       = {10.1126/science.1142188},
  publisher = {American Association for the Advancement of Science (AAAS)},
}

@Article{Lorenzoni2025,
  author    = {Lorenzoni, Nicola and Lacroix, Thibaut and Lim, James and Tamascelli, Dario and Huelga, Susana F. and Plenio, Martin B.},
  journal   = {Sci. Adv.},
  title     = {Full microscopic simulations uncover persistent quantum effects in primary photosynthesis},
  year      = {2025},
  issn      = {2375-2548},
  month     = oct,
  number    = {40},
  volume    = {11},
  doi       = {10.1126/sciadv.ady6751},
  fjournal  = {Science Advances},
  publisher = {American Association for the Advancement of Science (AAAS)},
}

\appendix
% ===================== Appendix A =====================
\section{End Matter}

\paragraph{Appendix A: Open quantum system model}
\label{app:model_class}
-To model an FMO dimer interacting with its environment under optical control, we consider a driven excitonic system linearly coupled to a Gaussian bosonic bath. This minimal open-system setting is chosen as a surrogate to isolate the essential ingredients responsible for correlation transfer and \textbf{nonsecular population–coherence transfer} in two-dimensional electronic spectroscopy (2DES).

The total Hamiltonian is written as
\begin{equation}
H(t)=H_S + H_B + H_{SB} + H_{\mathrm{int}}(t),
\label{eq:total_H}
\end{equation}
where $H_B$ is a bath of harmonic oscillators and $H_{SB}$ is the system--bath coupling.
The light--matter interaction $H_{\mathrm{int}}(t)$ is specified when defining the pulse unitaries.

%\paragraph*{FMO dimer and spectroscopy Hilbert space.}
We employ a minimal two-site Frenkel exciton model for the \emph{site-1/site-3} channel in the FMO complex dimer.
This choice is directly motivated by the original 2DES observations of long-lived beatings reported by Engel \emph{et al.}, which were first discussed in connection with the site-1/site-3 channel~\cite{Engel2007}.
In the site basis we use the four-level notation
$\{\ket{g},\ket{\epsilon_1},\ket{\epsilon_2},\ket{f}\}$,
where $\ket{\epsilon_1}$ and $\ket{\epsilon_2}$ denote local single-exciton states on site~1 and site~3, respectively.
The system Hamiltonian takes the standard Frenkel form,
\begin{equation}
\begin{aligned}
H_S
&=
\sum_{m=1}^{2}\epsilon_m \ket{\epsilon_m}\!\bra{\epsilon_m}
+
J\left(\ket{\epsilon_1}\!\bra{\epsilon_2}+\ket{\epsilon_2}\!\bra{\epsilon_1}\right) \\
&\quad+
\left(\epsilon_1+\epsilon_2\right)\ket{f}\!\bra{f}.
\end{aligned}
\label{eq:HS_site4}
\end{equation}

Diagonalizing the single-exciton block of Eq.~\eqref{eq:HS_site4} defines exciton eigenstates $\ket{1},\ket{2}$ with energies $E_1,E_2$.
We then use the exciton-basis notation $\{\ket{0},\ket{1},\ket{2},\ket{f}\}$ with $\ket{0}\equiv\ket{g}$, in which
\begin{equation}
H_S
=
E_1 \ket{1}\!\bra{1}
+
E_2 \ket{2}\!\bra{2}
+
\left(E_1+E_2\right)\ket{f}\!\bra{f}.
\label{eq:HS_exciton4}
\end{equation}

%\paragraph*{Bosonic bath and coupling operator.}
For FMO complexes, the relevant environment is the protein scaffold and surrounding solvent that interact with the pigments.
We model this environment as a bosonic bath that captures both fast fluctuations and slower collective motions, which together set the decoherence timescale and the degree of non-Markovian memory observed in 2DES. Two independent baths act at each sited, and both are modeled as reservoirs of harmonic oscillators with identical Hamiltonian,
\begin{equation}
H_B=\sum_k \omega_k b_k^\dagger b_k,
\end{equation}
and the system--bath coupling is taken in the linear (spin--boson) form
\begin{equation}
\begin{aligned}
    &H_{SB}=\ket{\epsilon_1} \bra{\epsilon_1}\otimes B_1 + \ket{\epsilon_2} \bra{\epsilon_2}\otimes B_2,
\qquad \\&
B_1=\sum_k g_{k1}\left(b_{k1}^\dagger+b_{k1}\right),\\&
B_2=\sum_k g_{k2}\left(b_{k2}^\dagger+b_{k2}\right).
\label{eq:HSB}
\end{aligned}
\end{equation}
As the optical transition rate is modest compared to the experimental time scale, we only track dissipative dynamics within the single-exciton manifold.

%\paragraph*{Spectral density and memory-bath regimes.}
The Gaussian bath is characterized by its spectral density,
\begin{equation}
J(\omega)\equiv \lambda^2\,\pi\sum_k g_k^2\,\delta(\omega-\omega_k),
\label{eq:J_def}
\end{equation}
or equivalently by the corresponding equilibrium correlation function.
For a broad class of memory baths we use the canonical power-law form with a cutoff,
\begin{equation}
J(\omega)=2\pi\,\lambda^2\,\omega_c^{\,1-s}\,\omega^{s}\,e^{-\omega/\omega_c},
\label{eq:J_powerlaw}
\end{equation}
where $\omega_c$ is the cutoff frequency and $s$ is the bath exponent.
The classification is: $s=1$ (Ohmic), $s<1$ (sub-Ohmic), and $s>1$ (super-Ohmic).
When $s<1$, the enhanced low-frequency weight leads to slower decay of bath correlations, realizing a memory bath over inter-pulse delays and waiting times.
We also consider structured low-frequency environments that similarly prolong bath memory. 

%\paragraph*{Structured low-frequency bath as an effective memory enhancement.}
To connect structured environments to microscopic vibrational degrees of freedom, we model a \emph{structured} low-frequency bath as an effective representation of a collection of low-frequency vibrational modes coupled to the excitonic dimer. Low-frequency vibrational motions are commonly observed in experiments on pigment--protein complexes~\cite{Renger2012,Klinger2020}.
A small subset of such modes, when concentrated near $\omega\approx 0$ or distributed with enhanced low-frequency weight, can be modeled as a modified spectral density that prolongs bath correlations and thereby increases the effective memory time.
Guided by this picture, we implement a minimal modification of the canonical Ohmic spectral density by adding a weak low-frequency tail near zero frequency, yielding a compact structured-bath model that mimics the slow-memory condition required for non-secular population–coherence transfer.

Specifically, we supplement the Ohmic form with a weak linear low-frequency tail,
\begin{equation}
J(\omega)=J_{\mathrm{Ohmic}}(\omega)+0.01\,J_{\mathrm{lin}}(\omega),
\label{eq:J_structured_def}
\end{equation}
where $J_{\mathrm{Ohmic}}(\omega)$ is the baseline Ohmic spectral density used above, and
\begin{equation}
J_{\mathrm{lin}}(\omega)=2\pi\,\lambda^2\,
\left(\frac{\omega}{\omega_c}\right)\,\Theta(0.05\,\omega_c-\omega).
\label{eq:J_lin_tail}
\end{equation}
Here $\Theta(\cdot)$ is the Heaviside function and the tail is restricted to the low-frequency window $\omega<0.05\,\omega_c$.
Although the added component is small in overall weight, it serves two purposes: 
1. selectively enhancing the low-frequency sector of \(J(\omega)\), and 2. introducing a strong structured element through the discontinuity at \(0.05 \omega_c\).
These effects directly contribute to slower bath correlation decay, resulting in an extended effective bath memory time, \emph{without} assuming a sub-Ohmic bath exponent.
% ---------- Add at the end of the Open quantum system model section ----------
%\paragraph*{Bloch--Redfield master equation, dissipator, and unitary dressing.}
Throughout we propagate the reduced state $\varrho(t)$ in the rotating-frame using a time-dependent Bloch--Redfield master equation that retains non-secular transfer~\cite{Chen2025}.
Let $U_S(t,\tau)$ denote the system propagator generated by the driven system Hamiltonian $H_S+H_{\mathrm{int}}(t)$, and let
$C(t)\equiv \mathrm{Tr}_B\!\left[\rho_B\,B(t)\,B(0)\right]$
be the bath correlation function.
The rotating frame Bloch--Redfield \emph{dissipator operators} are defined as
\begin{equation}
\Lambda(t)
= e^{iH_St}
\int_{0}^{t} d\tau\;
C(t-\tau)\;
U_S(t,\tau)\,A_0\,U_S(\tau,t)\,e^{-iH_St},
\label{eq:Lambda_def_IP}
\end{equation}
which enter the Bloch--Redfield equation in rotating-frame as
\begin{equation}
\frac{d}{dt}\varrho(t)
=
\big[\Lambda(t)\,\varrho(t),A_0\big]
+
\big[A_0,\varrho(t)\,\Lambda^\dagger(t)\big]
.
\label{eq:BR_ME_main}
\end{equation}
\(A_0\) is the coupling operator, which can be obtained via the method mentioned in Appendix C~\ref{app: frenkel2qubit}.
For numerical propagation it is convenient to vectorize the density operator, $|\varrho(t)\rangle\!\rangle \equiv \mathrm{vec}[\varrho(t)]$,
so that $d|\varrho(t)\rangle\!\rangle/dt = \mathcal{D}(t)\,|\varrho(t)\rangle\!\rangle$ with a dissipative generator $\mathcal{D}(t)$.
The corresponding \emph{dissipative generator} is the Liouville-space representation of Eq.~\eqref{eq:BR_ME_main}:
\begin{equation}
\mathcal{D}(t)
=
\Lambda^*(t)\otimes A_0
+
A_0^*\otimes \Lambda(t)
-
\mathbb{I}\otimes  A_0\,\Lambda(t)
-
A_0^*\,\Lambda^*(t)\otimes \mathbb{I}.
\label{eq:D_from_Lambda}
\end{equation}

In our notation, $\Lambda_S(t)$ denotes this static (factorized) non-Markovian dissipator within Bloch–Redfield theory and can be directly obtained by setting \(U_S(t,\tau)\) to be field free propagator in Eq.~\ref{eq:Lambda_def_IP}. It retains a finite degree of non-Markovianity associated with the initial transient during which correlations build up between the system and the bath following excitation. However, by construction, it neglects any system–bath correlations that exist \emph{prior} to the application of the pulse. The corresponding static Markovian dissipator, $\Lambda^{\mathrm{SM}}$, is defined as the long-time, time-independent limit ~\cite{Chen2025},
\begin{equation}
\Lambda^{\mathrm{SM}} = \lim_{t \to \infty} \Lambda_S(t).
\end{equation}

Since femtosecond pulses act on timescales much shorter than the bath correlation time, neither the assumption of a factorized initial state nor that of an equilibrated bath is justified. {\it In this regime, the dissipator itself must be propagated through the pulse.} Our key finding in Ref.~\cite{Chen2025} was that the dissipator which correctly accounts for pre-existing system–bath correlations prior to the pulse is given by a simple superposition of $\Lambda_S(t)$ and $\Lambda^{\mathrm{SM}}$. Following the pulse, the bath retains memory of the previously correlated state through the term $\Lambda^{\mathrm{SM}} - \Lambda_S(t)$, which relaxes only slowly as the bath reorganizes around the post-pulse excitonic configuration. The resulting time-dependent dissipator takes the form
\begin{equation}
\Lambda(t) = \Lambda_S(t) + U_1(t)\big(\Lambda^{\mathrm{SM}} - \Lambda_S(t)\big)U_1^\dagger(t),
\end{equation}
leading to a decomposition $\mathcal{D}(t) = \mathcal{D}_S(t) + \mathcal{D}_{\mathrm{mem}}(t)$.

Here, $U_1$ denotes the unitary pulse operator that promotes the system from the ground state to the excitonic manifold. Crucially, this pulse \emph{dresses the bath memory} by rotating population–population, population–coherence, coherence–population, and coherence–coherence sectors of the dissipator, with the resulting structure determined by the specific pulse unitary. As a consequence, averaging the post-pulse dynamics over timescales long compared to the excitonic splitting but short compared to the bath reorganization time does \emph{not} eliminate nonsecular contributions, in stark contrast to both factorized non-Markovian and Markovian treatments. Standard approaches that do not propagate dissipative structures across the pulse unitary therefore fail to capture the persistent population–coherence transfer responsible for the long-lived coherences observed in our model.

A schematic illustration of the correlation-aware dissipator construction and its contrast with conventional factorized treatments is shown in Fig.~\ref{fig:scheme_nonsecular}.

Finally, \emph{dynamical preparation} refers to the fact that the system-bath correlations are established before the pulse and carried through the pulse via Eq.~\eqref{eq:Lambda_def_IP}.
When bath correlations persist across inter-pulse delays, this unitary action can redistribute pre-existing system--bath correlations into
post-pulse system coherences, thereby modifying which dissipative pathways are active during subsequent evolution.

\paragraph{Appendix B: Explicit pulse-dressed dissipators for a three-pulse 2DES protocol}
\label{app:dissipator_explicit}

-We denote the three interpulse delays in a three-pulse 2DES protocol by $(t_1,t_2,t_3)$,
corresponding to the coherence time, waiting time, and detection time, respectively.
The three optical pulses are modeled as instantaneous system unitaries $U_{c1}$, $U_{c2}$, and $U_{c3}$, while $H_0$ is the field-free system Hamiltonian. Throughout the main text and numerics we take $U_{c1}=U_{c2}=U_{c3}\equiv U_c$. $H_0$ is equivalently $H_S$ in the main text.

%\paragraph*{Rotating frame Bloch--Redfield form and static dissipators.}
In the rotating frame with respect to $H_0$, we write the reduced density operator as
$\varrho(t)=e^{+iH_0 t}\rho(t)e^{-iH_0 t}$.
For a linear system--bath coupling $H_{SB}= A_0\otimes B$ and a Gaussian bath,
the Bloch--Redfield equation can be written in the compact commutator form
\begin{equation}
\frac{d}{dt}\varrho(t)
=
\big[\Lambda(t)\,\varrho(t),A_0\big]
+
\big[A_0,\varrho(t)\,\Lambda^\dagger(t)\big],
\label{eq:appB_BR_IP_general}
\end{equation}
where the dissipator $\Lambda(t)$ is discussed in main text.

In the \emph{conventional factorized} treatment of pulsed spectroscopy, dissipation during each field-free interval
is typically described using a ``static'' dissipator built under the assumption that the system--bath state
is factorized at the pulse boundary, with the bath reset to thermal equilibrium.
In this language, $\Lambda_S(t)$ denotes a \emph{static non-Markovian} (time-dependent) dissipator for the
field-free Hamiltonian $H_0$ . It retains partial memory over a time $t$ since the factorization,
and $\Lambda^{\mathrm{SM}}$ denotes the corresponding \emph{static Markovian} long-time limit,
\begin{equation}
\Lambda^{\mathrm{SM}} \equiv \lim_{t\to\infty}\Lambda_S(t),
\label{eq:appB_LSM_def}
\end{equation}
which becomes time-independent when bath correlations decay sufficiently rapidly.
Upon applying an additional secular approximation one recovers the widely used GKSL/Lindblad form.

%\paragraph*{Pulse-dressed unitaries.}
To describe how residual memory contributions are transported across pulse boundaries in our correlation-aware
construction, we introduce pulse-dressed unitaries (in the Schr\"odinger picture) that rotate the operator
content of the memory terms.

First, for the coherence-time segment 
\begin{equation}
U_1(t_1)
= e^{-i H_0 t_1}\,U_{c1}\,e^{+i H_0 t_1}.
\label{eq:appB_U1_full}
\end{equation}
For the waiting-time segment 
\begin{equation}
U_2(t_1,t_2)
= e^{-i H_0 t_2}\,U_{c2}\,e^{-i H_0 t_1}\,U_{c1}\,e^{+i H_0 t_1}\,e^{+i H_0 t_2}.
\label{eq:appB_U2_full}
\end{equation}
For the detection-time segment 
\begin{equation}
\begin{aligned}
U_3(t_1,t_2,t_3)
&= e^{-i H_0 t_3}\,U_{c3}\,e^{-i H_0 t_2}\,U_{c2}\,e^{-i H_0 t_1}\,U_{c1} \\
&\qquad \times e^{+i H_0 t_1}\,e^{+i H_0 t_2}\,e^{+i H_0 t_3}.
\end{aligned}
\label{eq:appB_U3_full}
\end{equation}

%\paragraph*{Explicit dissipators for the three segments.}
Following derivation in Ref.~\cite{Chen2025}, with the conventions above, the effective dissipators entering the three segments of the protocol can be written
in terms of the static dissipators $\Lambda_S(t)$ and $\Lambda^{\mathrm{SM}}$ as follows.

\emph{Coherence-time segment ($t_1$):}
\begin{equation}
\Lambda_1(t_1)
=
U_1(t_1)\Big[\Lambda^{\mathrm{SM}}-\Lambda_S(t_1)\Big]U_1^\dagger(t_1).
\label{eq:appB_L1_full}
\end{equation}

\emph{Waiting-time segment ($t_2$):}
\begin{align}
&\Lambda_2(t_1,t_2)
=
U_2(t_1,t_2)\Big[\Lambda^{\mathrm{SM}}-\Lambda_S(t_1+t_2)\Big]U_2^\dagger(t_1,t_2)
\nonumber\\
&\quad
+\,
U_1(t_2)\Big[\Lambda_S(t_1+t_2)-\Lambda_S(t_2)\Big]U_1^\dagger(t_2)
+\Lambda_S(t_2).
\label{eq:appB_L2_full}
\end{align}

\emph{Detection-time segment ($t_3$):}
\begin{align}
&\Lambda_3(t_1,t_2,t_3)
=\\&
U_3(t_1,t_2,t_3)\Big[\Lambda^{\mathrm{SM}}-\Lambda_S(t_1+t_2+t_3)\Big]U_3^\dagger(t_1,t_2,t_3)
\nonumber\\
&\quad
+\,
U_2(t_1,t_3)\Big[\Lambda_S(t_1+t_2+t_3)-\Lambda_S(t_2+t_3)\Big]U_2^\dagger(t_1,t_3)
\nonumber\\
&\quad
+\,
U_1(t_3)\Big[\Lambda_S(t_2+t_3)-\Lambda_S(t_3)\Big]U_1^\dagger(t_3)
+\Lambda_S(t_3).
\label{eq:appB_L3_full}
\end{align}

Equations~\eqref{eq:appB_L1_full}--\eqref{eq:appB_L3_full} make explicit that the pulse sequence
\emph{unitarily dresses} the residual memory contributions through $U_1$, $U_2$, and $U_3$. The resulting dissipators are substituted into Eq.~\eqref{eq:appB_BR_IP_general} to propagate the reduced dynamics.

\paragraph{Appendix C: Third-order response construction for 2DES}
\label{app:third_order_2des}

%\paragraph*{Third-order signal measured in 2DES.}
-In a three-pulse 2DES experiment, the detected field is, to leading nonlinear order, a \emph{third-order} response of an open quantum system to a prescribed sequence of ultrafast light--matter interactions~\cite{Biswas2022,Chen2010}.
The signal is recorded in the time domain as a function of the coherence time $t_1$, waiting time $T$, and detection time $t_3$, and then converted to a two-frequency spectrum by Fourier transforms along $t_1$ and $t_3$.

%\paragraph*{Response-function construction and its limitation for non secular dynamics.}
A standard impulsive-response representation expresses the rephasing and nonrephasing third-order response functions as
\begin{align}
R_{\mathrm{rp}}(t_3,T,t_1)
&=
\big\langle\!\big\langle
\mu_{-}\,\mathcal{G}(t_3)\,V_{+}\,\mathcal{G}(T)\,V_{+}\,\mathcal{G}(t_1)\,V_{-}\,\rho_0
\big\rangle\!\big\rangle,
\label{eq:Rrp_def}
\\
R_{\mathrm{nr}}(t_3,T,t_1)
&=
\big\langle\!\big\langle
\mu_{-}\,\mathcal{G}(t_3)\,V_{+}\,\mathcal{G}(T)\,V_{-}\,\mathcal{G}(t_1)\,V_{+}\,\rho_0
\big\rangle\!\big\rangle,
\label{eq:Rnr_def}
\end{align}
where $V\rho\equiv[\mu,\rho]$ is the commutator superoperator and $\mathcal{G}(t)$ denotes the open-system propagation between optical interactions.
In many conventional treatments, these expressions are evaluated using propagators for which population and coherence sectors are effectively decoupled during the field-free intervals, enabling an intuitive organization in terms of Liouville pathways~\cite{Chen2010,Biswas2022}.

In our correlation-aware dynamics, the field-free evolution is governed by a pulse-dressed dissipative generator that can exhibit \textbf{non-secular population to coherence transfer} during the waiting-time evolution.
This population--coherence mixing is an essential ingredient of the long-lived beatings in our simulations.
Under such dynamics, however, a direct use of Eqs.~\eqref{eq:Rrp_def}--\eqref{eq:Rnr_def} as the primary computational definition can can obscure the fact that the non-secular population–coherence transfer arises dynamically during field-free evolution, rather than from a fixed Liouville-pathway partitioning.

%\paragraph*{Operational third-order extraction by inclusion--exclusion.}
Under the weak-field approximation, the net action of an ultrashort optical interaction can be represented as a small unitary rotation generated by the transition dipole,
\begin{equation}
\mathcal{U}(\epsilon)\rho \equiv U(\epsilon)\rho U^\dagger(\epsilon),
\qquad
U(\epsilon)=\exp(-i\epsilon\,\mu),
\label{eq:Ueps_def}
\end{equation}
so that
\begin{equation}
\begin{aligned}
    \mathcal{U}(\epsilon)\rho
=
\rho - i\epsilon[\mu,\rho] + &\mathcal{O}(\epsilon^2)
=
\rho - i\epsilon\,V\rho + \mathcal{O}(\epsilon^2),
\qquad\\&
V\rho\equiv[\mu,\rho].
\label{eq:Ueps_linear}
\end{aligned}
\end{equation}
The 2DES signal is the third-order contribution in $\epsilon$, i.e., the component that requires three optical interactions~\cite{Biswas2022}.

To obtain an \emph{equivalent} third-order signal while ensuring that \textbf{non-secular population to coherence transfer} contained in our correlation-aware propagation is taken into account, we extract the third-order component operationally from the full driven open-system dynamics.
We compute a set of time-domain signals in which each of the three optical interactions is either included or excluded at the level of the applied optical action, while keeping the same open-system propagation (and thus the same non-secular effects) throughout the protocol.

To make this precise, we introduce a binary ``switch'' for the optical-action superoperators:
\begin{equation}
\mathcal{V}^{(a)}_{\pm} \equiv \mathbb{I}-\,ia\epsilon\,V_{\pm},
\qquad a\in\{0,1\},
\label{eq:V_switch}
\end{equation}
where $\mathbb{I}$ is the identity superoperator on the system density operator.
We then define the phase-selected rephasing/nonrephasing time-domain signals with switches $(a,b,c)\in\{0,1\}^3$ as
\begin{align}
S^{\mathrm{rp}}_{abc}(t_3,T,t_1)
&\equiv
\big\langle\!\big\langle
\mu_{-}\,\mathcal{G}(t_3)\,\mathcal{V}^{(c)}_{+}\,\mathcal{G}(T)\,\mathcal{V}^{(b)}_{+}\,\mathcal{G}(t_1)\,\mathcal{V}^{(a)}_{-}\,\rho_0
\big\rangle\!\big\rangle,
\label{eq:Sabc_rp}
\\
S^{\mathrm{nr}}_{abc}(t_3,T,t_1)
&\equiv
\big\langle\!\big\langle
\mu_{-}\,\mathcal{G}(t_3)\,\mathcal{V}^{(c)}_{+}\,\mathcal{G}(T)\,\mathcal{V}^{(b)}_{-}\,\mathcal{G}(t_1)\,\mathcal{V}^{(a)}_{+}\,\rho_0
\big\rangle\!\big\rangle.
\label{eq:Sabc_nr}
\end{align}
The pure third-order component is then isolated by inclusion--exclusion,
\begin{equation}
\begin{aligned}
S^{(3)}_{\mathrm{R/NR}}(t_3,T,t_1)
&=(-i)^3(
S^{\mathrm{R/NR}}_{111}-S^{\mathrm{R/NR}}_{110}-\\&S^{\mathrm{R/NR}}_{101}-S^{\mathrm{R/NR}}_{011}
+S^{\mathrm{R/NR}}_{100}\\&+S^{\mathrm{R/NR}}_{010}+S^{\mathrm{R/NR}}_{001}-S^{\mathrm{R/NR}}_{000}).
\label{eq:IE_3rd}
\end{aligned}
\end{equation}
Equation~\eqref{eq:IE_3rd} cancels 0th-, 1st-, and 2nd-order contributions by construction, leaving the component that requires all three optical interactions.
Because the underlying propagation is the full correlation-aware evolution introduced in the previous section, the resulting $S^{(3)}_{\mathrm{R/NR}}$ retains non-secular population to coherence transfer and provides the appropriate input for our 2DES analysis~\cite{Chen2025}.

%\paragraph*{2D spectra and observable.}
The corresponding 2D spectra are obtained by Fourier transform along $t_1$ and $t_3$,
\begin{equation}
\begin{aligned}
    \tilde S^{(3)}_{\mathrm{R/NR}}(\omega_3,T,\omega_1)
=
\int_{0}^{\infty}\!dt_3\int_{0}^{\infty}\!dt_1\;\\
e^{+i\omega_3 t_3}\,e^{\pm i\omega_1 t_1}\,
S^{(3)}_{\mathrm{R/NR}}(t_3,T,t_1),
\label{eq:FT_2D_3rd}
\end{aligned}
\end{equation}
and we report the absorptive spectrum as
\begin{equation}
S_{\mathrm{abs}}(\omega_3,T,\omega_1)
=
\mathrm{Re}\!\left[
\tilde S^{(3)}_{\mathrm{R}}(\omega_3,T,\omega_1)
+
\tilde S^{(3)}_{\mathrm{NR}}(\omega_3,T,\omega_1)
\right].
\label{eq:absorptive_3rd}
\end{equation}
To quantify long-lived beatings, we track the waiting-time dependence of a selected cross-peak amplitude in the absorptive spectrum. Beating amplitude and phase are extracted from $A_{\mathrm{CP}}(T)$ and processed by Fourier analysis to provide frequency domain information.

% ============================================================
% Appendix: Two local baths -> single effective coupling operator
% ============================================================
\paragraph{Appendix D: Mapping Frenkel exciton model to qubit model}\label{app: frenkel2qubit}

-We start from a minimal site-basis excitonic dimer coupled to two \emph{local} (independent) harmonic environments,
\begin{align}
H_{\mathrm{tot}}
&= H_S + H_B + H_{SB}, \label{eq:appC_Htot}
\\
H_S
&= \epsilon_1 \ket{1}\!\bra{1} + \epsilon_2 \ket{2}\!\bra{2}
+ J\big(\ket{1}\!\bra{2}+\ket{2}\!\bra{1}\big), \label{eq:appC_HS_site}
\\
H_B
&= H_{B1}+H_{B2},
\qquad
H_{Bi}=\sum_k \omega_k\, b_{k,i}^\dagger b_{k,i}, \label{eq:appC_HB}
\\
H_{SB}
&= \ket{1}\!\bra{1}\,\otimes B_1 \;+\; \ket{2}\!\bra{2}\,\otimes B_2, 
\qquad
\\B_i&=\sum_k g_k\big(b_{k,i}^\dagger+b_{k,i}\big),
\label{eq:appC_HSB_local}
\end{align}
where $\ket{1},\ket{2}$ denote the site-local single-excitation states, and the two baths are taken to be uncorrelated and
Gaussian.

We rewrite the site projectors as $\ket{1}\!\bra{1}=\tfrac12(\mathbb{I}+\sigma_z)$ and
$\ket{2}\!\bra{2}=\tfrac12(\mathbb{I}-\sigma_z)$ (in the single-excitation site subspace). This gives
\begin{equation}
H_{SB}
=\frac{\mathbb{I}}{2}\otimes(B_1+B_2)\;+\;\frac{\sigma_z}{2}\otimes(B_1-B_2).
\label{eq:appC_sumdiff_short}
\end{equation}
The first term is proportional to the identity on the system and therefore does not influence the reduced system
dynamics. We thus keep only the
\emph{difference-bath} coupling and define
\begin{equation}
B_{\Delta}\equiv \frac{B_1-B_2}{\sqrt{2}},
\qquad
H_{SB}\;\simeq\;\frac{\sigma_z}{\sqrt{2}}\otimes B_{\Delta}.
\label{eq:appC_diffbath_short}
\end{equation}

To express the coupling in the exciton eigenbasis, diagonalize the site Hamiltonian $H_S$ and denote by
$\{\ket{e_1},\ket{e_2}\}$ the eigenstates with energies $E_{e_1}<E_{e_2}$ and splitting $\Delta\equiv E_{e_2}-E_{e_1}$.
Let $U$ be the $2\times 2$ unitary that maps the site basis to the exciton basis,
$\ket{e_\alpha}=\sum_{i=1}^2 U_{i\alpha}\ket{i}$.
Then the system coupling operator in the exciton basis is obtained by the basis rotation
\begin{equation}
A_0 \;\equiv\; U^\dagger \sigma_z U.
\label{eq:appC_A_def}
\end{equation}
For a dimer Hamiltonian, $U$ can be written in terms of a mixing angle $\theta$ as
\begin{equation}
\begin{aligned}
    &\ket{e_1}=\cos\theta\,\ket{1}-\sin\theta\,\ket{2},
\qquad
\\&\ket{e_2}=\sin\theta\,\ket{1}+\cos\theta\,\ket{2},
\qquad
\\&\tan(2\theta)=\frac{2J}{\epsilon_1-\epsilon_2},
\label{eq:appC_theta_short}
\end{aligned}
\end{equation}
which yields the compact exciton-basis form
\begin{equation}
A_0
=
\cos(2\theta)\,\sigma_z^{(e)}+\sin(2\theta)\,\sigma_x^{(e)}.
\label{eq:appC_A_exciton}
\end{equation}
Equations~\eqref{eq:appC_diffbath_short}--\eqref{eq:appC_A_exciton} therefore show that two local baths reduce to a
single dynamically relevant difference bath coupled through the exciton-basis operator $A$,
\begin{equation}
H_{SB}\;\simeq\;\frac{1}{\sqrt{2}}\,A_0\otimes B_{\Delta},
\end{equation}
which is the coupling structure used throughout the main text.

\paragraph{Appendix E: Identifying inter-site entanglement in the FMO dimer}
\label{appendix: entanglement}

-Two distinct issues are routinely conflated in discussions of long-lived beatings: their \emph{origin} (purely excitonic dynamics in the presence of a non-flat environmental spectral density versus vibronic contributions) and their \emph{nature} (classical versus quantum). In a dimer mapped onto an effective two-level dissipative quantum system, the relevant nonclassical ingredient is most cleanly discussed at the level of \emph{coherence}: off-diagonal elements in the density matrix represent superposition and are therefore quantum mechanical in character. At the same time, in the four-level spectroscopy space this does \emph{not} imply quantum entanglement between the two sites. Entanglement certification via the Peres--Horodecki criterion requires coherence of sufficiently large magnitude relative to the associated populations for a chosen bipartition so that the partial transpose becomes non-positive, a condition that is typically not met under weak-field excitation in third-order spectroscopy~\cite{Peres1996,Horodecki1996}. The classical--quantum dichotomy becomes meaningful only in extended multisite systems, where one may separately analyze coherence and inter-site entanglement as distinct notions of nonclassicality.

The ``quantumness'' of long-lived beatings are often discussed in more than one sense. Two notions that are commonly considered are:
(i) \emph{coherence} in the reduced density operator, defined (in a chosen basis) by the presence of off-diagonal density-matrix elements; and
(ii) \emph{entanglement} between spatial subsystems, a stronger and genuinely multipartite form of nonclassical correlation that cannot be inferred from off-diagonal elements alone.

This Appendix summarizes the operational entanglement test used for the FMO dimer and clarifies how it relates to the coherences analyzed in the main text~\cite{Peres1996,Horodecki1996}.

%\paragraph*{Two-site bipartition and basis mapping.}
Modeling each pigment as a local two-level system yields a bipartite Hilbert space
$\mathcal{H}=\mathcal{H}_1\otimes\mathcal{H}_2$ with $\dim\mathcal{H}_1=\dim\mathcal{H}_2=2$.
In the site basis, we identify the spectroscopy basis $\{|g\rangle,|\epsilon_1\rangle,|\epsilon_2\rangle,|f\rangle\}$
with the two-qubit product basis $\{|00\rangle,|10\rangle,|01\rangle,|11\rangle\}$ as
\begin{equation}
|g\rangle \equiv |00\rangle,\qquad
|\epsilon_1\rangle \equiv |10\rangle,\qquad
|\epsilon_2\rangle \equiv |01\rangle,\qquad
|f\rangle \equiv |11\rangle .
\label{eq:basis_map_twoqubit_app}
\end{equation}
Inter-site entanglement is then defined with respect to the site bipartition $(1|2)$ in this product structure.

%\paragraph*{Peres--Horodecki criterion.}
For two qubits, separability across $(1|2)$ is equivalent to positivity of the partial transpose according to
Peres--Horodecki theorem~\cite{Peres1996,Horodecki1996}.
Writing matrix elements as $\varrho_{i\mu,j\nu}=\langle i\mu|\varrho|j\nu\rangle$,
the partial transpose with respect to site 2 is
\begin{equation}
\big(\varrho^{T_2}\big)_{i\mu,j\nu} \equiv \varrho_{i\nu,j\mu}.
\label{eq:PT_def_app}
\end{equation}
For two qubits, $\varrho$ is entangled if and only if $\varrho^{T_2}$ has at least one negative eigenvalue.

%\paragraph*{Illustrative example: threshold of a single-exciton coherence.}
To show whether coherence components can certify inter-site entanglement, consider the simplified
$4\times4$ density matrix 
\begin{equation}
\varrho \;=\;
\begin{pmatrix}
\rho_{gg} & 0 & 0 & 0\\
0 & \rho_{11} & e & 0\\
0 & e^{*} & \rho_{22} & 0\\
0 & 0 & 0 & \rho_{ff}
\end{pmatrix},
\end{equation}
where
\begin{equation}
\rho_{gg},\rho_{11},\rho_{22},\rho_{ff}\ge 0,\quad
\mathrm{Tr}\,\varrho=1,
\label{eq:X_state_example}
\end{equation}
where $e$ occupies the \emph{inter-site single-exciton coherence} position
($|10\rangle\langle 01|$ and its adjoint).
The partial transpose with respect to site 2 moves this coherence into the $\{|00\rangle,|11\rangle\}$ block,
\begin{equation}
\varrho^{T_2} \;=\;
\begin{pmatrix}
\rho_{gg} & 0 & 0 & e\\
0 & \rho_{11} & 0 & 0\\
0 & 0 & \rho_{22} & 0\\
e^{*} & 0 & 0 & \rho_{ff}
\end{pmatrix}.
\label{eq:X_state_PT}
\end{equation}
The spectrum of $\varrho^{T_2}$ consists of $\rho_{11}$, $\rho_{22}$, and the two eigenvalues of the
$2\times2$ block $\begin{pmatrix}\rho_{gg}&e\\ e^{*}&\rho_{ff}\end{pmatrix}$, namely
\begin{equation}
\lambda_{\pm}
=
\frac{\rho_{gg}+\rho_{ff}}{2}
\pm
\frac{1}{2}\sqrt{(\rho_{gg}-\rho_{ff})^{2}+4|e|^{2}}.
\label{eq:PT_eigs}
\end{equation}
Peres--Horodecki condition is violated precisely when the smaller eigenvalue $\lambda_{-}$ becomes negative, which occurs if
\begin{equation}
|e|^{2} \;>\; \rho_{gg}\,\rho_{ff}.
\label{eq:PPT_threshold_example}
\end{equation}
Equation~\eqref{eq:PPT_threshold_example} illustrates a general point: the presence of a
coherence term does not, by itself, guarantee entanglement. Peres--Horodecki condition violation depends on how that coherence
compares to the relevant populations. 

%\paragraph*{Application to reduced states in 2DES.}
Under pulsed excitation, the reduced state $\varrho(t)$ generally contains additional off-diagonal elements,
including optical coherences (e.g., $|g\rangle\langle\epsilon_i|$), so the simplified structure in
Eq.~\eqref{eq:X_state_example} need not apply.
Moreover, coherences discussed in the main text are often most naturally expressed in the exciton basis,
while inter-site entanglement is defined with respect to the site bipartition; the Peres--Horodecki test must therefore
be performed on $\varrho(t)$ represented in the site product basis of Eq.~\eqref{eq:basis_map_twoqubit_app}.
In practice, we evaluate $\varrho^{T_2}(t)$ at all times during the 2DES protocol and find that its eigenvalues
remains nonnegative. As a result, no inter-site entanglement is
certified in the parameter regime considered here.

\end{document}